\documentclass[fleqn,twoside]{article}
\usepackage{espcrc2}

\usepackage{graphicx}
\usepackage[figuresright]{rotating}

\newcommand{\AmS}{{\protect\the\textfont2
  A\kern-.1667em\lower.5ex\hbox{M}\kern-.125emS}}
\newcommand{\be}{\begin{equation}}
\newcommand{\ee}{\end{equation}}

\title{The helicity modulus in gauge field theories}

\author{Michele Vettorazzo
        \address[ZURICH]{Institute for Theoretical Physics, ETH Z\"{u}rich,
        CH-8093 Z\"{u}rich, Switzerland}\thanks{Talk given by Michele Vettorazzo} and
        Philippe de Forcrand \addressmark[ZURICH]
        \address[CERN]{CERN, Theory Division, CH-1211 Gen\`{e}ve 23, Switzerland}
        }
\begin{document}

\begin{abstract}
We consider the $4$d compact $U(1)$ theory with Wilson action and characterize its phase diagram using
the notion of electromagnetic flux, instead of the more usual magnetic monopole. Taking
inspiration from the flux picture, we consider the helicity modulus (h.m.) for this theory,
and show that it is an order parameter for the confinement deconfinement phase transition.
We extend the definition of the h.m. to an Abelian projected Yang-Mills theory, and discuss
its behavior in $SU(2)$.
\end{abstract}

\maketitle

\vspace{-0.2cm}
\section{Introduction}
\vspace{-0.2cm} Consider a theory characterized by the presence of a phase transition as
one parameter $\beta$ (the inverse temperature or some coupling of the theory) varies
through some particular value $\beta_c$. A usual way to describe the transition is provided
by the so-called \emph{order parameter}, defined as a function of $\beta$ which is zero in
one phase and non-zero in the other. The meaning of this definition is two-fold:
on one hand, such a function has certainly a point of non-analyticity, which
reflects the underlying non-analyticity of the free energy of the system (in the
thermodynamic limit); on the other hand, this behavior can sometimes be used to relate the
phase transition with the spontaneous breaking of some symmetry (implicitly or explicitly
defined) of the model, thus providing an appealing physical picture of the transition
itself.

In this paper we introduce an order parameter for the $4d$ compact $U(1)$ Abelian theory,
and study it numerically for the Wilson action

\vspace{-0.15cm} \be S=-\beta \sum_P \cos \theta_P \ee\vspace{-0.2cm}

\noindent where $\beta$ is the inverse bare coupling and $\theta_P$ is the plaquette angle.
This theory is characterized by a strong coupling confined phase and a weak coupling
Coulomb phase; in the confined phase, in analogy with ordinary superconductors, the $U(1)$
gauge symmetry is supposed to be spontaneously broken. Our order parameter will make use
exclusively of the notion of electromagnetic flux (instead than the more usual magnetic
monopole) which we now introduce.

\vspace{-0.2cm}
\section{The flux in Abelian theories}
\vspace{-0.2cm}

Let $L$ be the lattice size of a $4d$ hypercubic lattice with periodic boundary conditions
(p.b.c.). Our definition of the flux $\Phi_{\mu\nu}$ through a given $(\mu,\nu)$ orientation is:
\vspace{-0.2cm}

\begin{equation}
\label{eq:lattice_flux}
\Phi_{\mu\nu}=\frac{1}{L^2}\sum_{\hspace{-0.15cm}\mbox{\tiny{$\begin{array}[2pt]{c}\mbox{\tiny{$(\mu,\nu)$}}
\\\mbox{\tiny{{\rm planes}}}\end{array}$}}}
\sum_{P_{\mu\nu}}\hspace{0.05cm}[(\theta_P)_{\mu\nu}]_{-\pi,\pi}
\end{equation}\vspace{-0.1cm}

\noindent where $[\theta_P]_{-\pi,\pi}$ is the plaquette angle reduced to the interval $[-\pi,\pi]$.
A double sum is present:

\vspace{-0.2cm}
\begin{itemize}
\item   the \emph{internal} $\sum_{P_{\mu\nu}}$ is the sum over the plaquettes in a single
plane, and is $2\pi k$ ($k \in Z$) valued because of the  p.b.c. .\vspace{-0.1cm}

\item  the \emph{external} average $\frac{1}{L^2}\sum_{\mu\nu \hspace{0.1cm} planes}$, over
all parallel planes of the given orientation, is non-trivial because the flux through
different planes can change due to the presence of magnetic monopoles (herein lies the
connection with the usual picture).
The allowed values for $\Phi_{\mu\nu}$ are thus multiples of $2\pi/L^2$.
\end{itemize}
\vspace{-0.2cm}

Monitoring the \emph{flux distribution} $\nu(\phi)$ in the two phases, one observes that
$\nu(\phi)$ is Gaussian (centered at $\phi=0$) in the confined phase, while it is peaked
around multiples of $2\pi$ in the Coulomb phase (defining so-called \emph{flux sectors});
in the thermodynamic limit, \emph{tunnelling} between flux sectors becomes completely
suppressed. This behavior already provides a characterization of the phase transition.
\vspace{-0.2cm}

\section{Response to an external flux}
\vspace{-0.2cm}

One can now ask what is the response of the system to an external
electromagnetic flux, in the two phases.
The corresponding {\em flux} free energy is a straightforward extension to the
Abelian case of 't Hooft's non-Abelian {\em twist} free energy \cite{'tHooft:1979uj}.
Like the latter \cite{DeForcrand:2001dp}, it has a characteristic behavior in each phase.
We impose an extra flux $\phi \in R$ to the
following stack of plaquettes (this is only one possible choice):

\vspace{-0.2cm} \be \label{eq:twisted_stack}{\rm stack}=\lbrace  \theta_{P_{\mu\nu}} \mid
\mu=1,\nu=2; x=1,y=1 \rbrace. \ee \vspace{-0.2cm}

\noindent The partition function of the system becomes

\vspace{-0.2cm} \be Z(\phi)=\int D\theta\hspace{0.05cm} e^{\beta(\sum_{\rm stack}\cos
(\theta_P+\phi)+ \sum_{\overline{\rm stack}} \cos \theta_P)}\ee\vspace{-0.1cm}

\noindent where `$\overline{\rm stack}$' is the complement of the stack, i.e., consists of
all the other unchanged plaquettes. Alternatively (and more conveniently from a numerical
point of view), it is possible to perform a change of variables such that the extra flux is
spread through all the plaquettes with the given orientation $(\mu,\nu)$, leading to

\vspace{-0.2cm} \be
\label{eq:extra_flux_spread}Z(\phi)\hspace{-0.08cm}=\hspace{-0.15cm}\int\hspace{-0.1cm}
D\theta\hspace{0.1cm}
e^{\hspace{0.1cm}\beta(\sum_{\hspace{-0.15cm}\mbox{\tiny{$\begin{array}[2pt]{l}
\mbox{\tiny{$(\mu,\nu)$}}\\\mbox{\tiny{{\rm planes}}}\end{array}$}}}\hspace{-0.25cm}\cos
(\theta_P+\frac{\phi}{L^2})+\sum_{\hspace{-0.15cm}\mbox{\tiny{$\begin{array}[2pt]{c}\mbox{\tiny{$\overline{(\mu,\nu)}$}}
\\\mbox{\tiny{{\rm planes}}}\end{array}$}}}\hspace{-0.15cm}\cos\theta_P)}\ee\vspace{-0.1cm}

\noindent where `$\overline{(\mu,\nu)}$' indicates all the other orientations, through
which no extra flux is imposed.

\vspace{-0.cm}
\begin{figure}[t]
\begin{center}
\includegraphics[angle=-90,width=6.0cm]{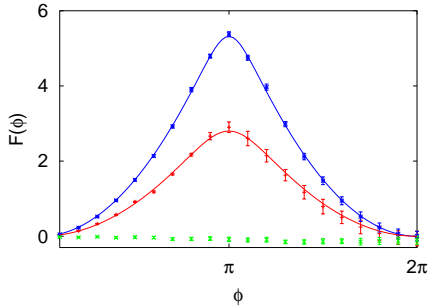}\end{center}
\vspace{-1.2cm} \caption{Flux free energy vs. the external flux $\phi$ in the Coulomb phase
(upper and middle curve, resp. $\beta=1.5, 1.1$) and in the confined phase (lower curve,
$\beta=0.8$). The fit using Eq.(\ref{eq:free_energy_ansatz}) is superimposed
($\beta_{\rm{R}}=1.2, 0.71$). The volume is $4^4$.} \label{fig:coulomb_confined}
\vspace{-0.7cm}
\end{figure}

%
%

$Z(\phi)$ is $2\pi$ periodic in $\phi$. Fig.\ref{fig:coulomb_confined} shows the behavior
of the free energy $F(\phi)=-\log Z(\phi)$ of the system in the two phases. In the confined
phase (where the flux can freely change) the system is insensitive to the presence of an
external flux ($F(\phi)={\rm const.}$). In the Coulomb phase the free energy is described
perfectly by the ansatz

\vspace{-0.2cm} \be\label{eq:free_energy_ansatz} F(\phi)=-\log \sum_k e^{-\frac{\beta_{\rm
R}}{2}(\phi-2\pi k)^2}\ee\vspace{-0.2cm}

\noindent which says that around each flux-sector the dependence on the external flux is
quadratic, as expected classically, and that we must consider the contribution of all
the sectors at the same time. $\beta_{\rm R}$ is the only parameter in this equation.
It replaces $\beta$ in the classical expression (\ref{eq:free_energy_ansatz}), and
therefore plays the role of a \emph{renormalized coupling}
\cite{Forcrand}.

\vspace{-0.2cm}
\section{The helicity modulus}\vspace{-0.2cm}

The last step in the construction of the order parameter is to note that the physical
information contained in Fig.\ref{fig:coulomb_confined} can be rephrased in a more concise
way: if instead of the whole curve (as a function of $\phi$) we consider only the
\emph{curvature} of $F(\phi)$ at the origin (or at any other point), we get a function of
$\beta$  which is always zero in the confined phase ($F(\phi)={\rm const.}$) and is
different from zero in the Coulomb phase; that is, it is an order parameter. This
construction was already known in the context of the $2d$ XY model \cite{Nelson}, where the
name  `helicity modulus' was first introduced. In our context we define the helicity modulus

\vspace{-0.1cm} \be\label{eq:hm_def} h(\beta)=\frac{\partial^2 F(\phi)}{\partial
\phi^2}\mid_{\phi=0} \ee\vspace{-0.1cm}

\noindent
which can be related to $\beta_{\rm R}$ via Eq.(\ref{eq:free_energy_ansatz}).
Computing explicitly the double derivative (with $F(\phi)$ defined as per
Eq.(\ref{eq:extra_flux_spread})), one gets

\vspace{-0.1cm} \be \label{eq:helicity_modulus_explicit}
h(\beta)\hspace{-0.05cm}=\hspace{-0.05cm}\frac{1}{V}\langle\hspace{-0.1cm}
\sum_{\hspace{-0.15cm}\mbox{\tiny{$\begin{array}[2pt]{c}\mbox{\tiny{$(\mu,\nu)$}}\\\mbox{\tiny{{\rm
planes}}}\end{array}$}}}\hspace{-0.15cm}\beta\cos\theta_P\rangle
-\frac{1}{V}\langle(\hspace{-0.15cm}\sum_{\hspace{-0.15cm}\mbox{\tiny{$\begin{array}[2pt]{l}
\mbox{\tiny{$(\mu,\nu)$}}\\\mbox{\tiny{{\rm
planes}}}\end{array}$}}}\hspace{-0.15cm}\beta\sin\theta_P)^2\rangle \ee \vspace{-0.2cm}

\begin{figure}[t]
\begin{center}\includegraphics[angle=-90,width=6.0cm]{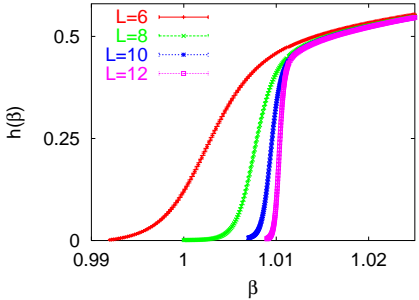}\end{center}
\vspace{-1.2cm} \caption{The helicity modulus vs. $\beta$.}
\label{fig:hm_rew_8p4_10p4_12p4_14p4} \vspace{-0.5cm}
\end{figure}

In Fig.\ref{fig:hm_rew_8p4_10p4_12p4_14p4} we plot this observable for different volumes:
its behavior is qualitatively consistent with an order parameter.
Moreover, it is shown in \cite{Forcrand} that the drop to zero in the confined phase is
exponential in $(\beta_c - \beta)$, providing convincing evidence that the transition is
$1^{\rm st}$ order.

\vspace{-0.2cm}
\section{Extension to non-Abelian theories}
\vspace{-0.2cm} We now turn our attention to Yang-Mills theories, which display
a finite temperature transition (say, at $T_c$) between a confining and a deconfined phase;
here we consider the gauge group $SU(2)$.

If we assume that (1), following \cite{'tHooft:1981ht}, the long range properties of the
theory can be well described by an effective Abelian theory, then we expect that the
Abelian gauge ensemble obtained after a suitable gauge fixing and Abelian projection must
change from confining to deconfined across $T_c$. Therefore, the helicity modulus measured
in the projected ensemble should present a discontinuity at $T_c$. Moreover, under the
further assumption (2) that the effective action which describes the projected ensemble is
Wilson-like

\vspace{-0.1cm}
 \be \label{eq:effective_action}S_{\rm eff}=\beta_{\rm eff}\sum_P \cos \theta_{P
\hspace{0.02cm}{\rm proj.}}\ee \vspace{-0.2cm}

\noindent where $\beta_{\rm eff}$ is an effective coupling whose value could be determined
by Inverse Monte Carlo, the helicity modulus is the same as in
Eq.(\ref{eq:helicity_modulus_explicit}), which we rewrite for convenience as

\vspace{-0.2cm} \be \label{eq:hel_mod_non_abelian} h(\beta)=\beta_{\rm eff} H_1(\beta)
-\beta_{\rm eff}^2 H_2(\beta) \vspace{-0.2cm}\ee\vspace{-0.4cm}

\be
\label{eq:H1}H_1(\beta)=\frac{1}{V}\langle\sum_{\hspace{-0.15cm}\mbox{\tiny{$\begin{array}[2pt]{l}
\mbox{\tiny{$(\mu,\nu)$}}\\\mbox{\tiny{{\rm planes}}}\end{array}$}}}\cos\theta_P\rangle
\vspace{-0.4cm}\ee\vspace{-0.4cm}

\be \label{eq:H2} \vspace{-0.2cm}
H_2(\beta)=\frac{1}{V}\langle(\sum_{\hspace{-0.15cm}\mbox{\tiny{$\begin{array}[2pt]{l}
\mbox{\tiny{$(\mu,\nu)$}}\\\mbox{\tiny{{\rm
planes}}}\end{array}$}}}\sin\theta_P)^2\rangle.\ee\vspace{-0.2cm}

To perform the Abelian projection, we use the Maximal Abelian Gauge \cite{Kronfeld:1987ri},
defined by

\begin{figure}[t]
\begin{center}\includegraphics[angle=-90,width=6.0cm]{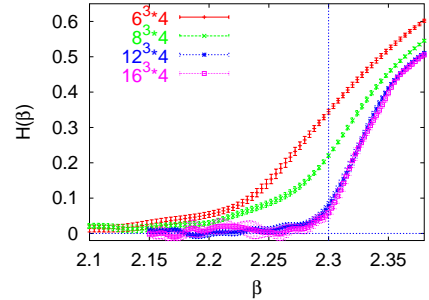}\end{center} \vspace{-1.2cm}
\caption{The helicity modulus Eq.(\ref{eq:hel_mod_non_abelian}) (non-Abelian case) for a
test value $\beta_{\rm eff}=1.0$.} \label{fig:hm_Mag_b_Beff1}\vspace{-0.5cm}
\end{figure}

\vspace{-0.2cm} \be
\sum_{x,\mu} Tr (U_\mu^\dagger(x)\sigma_3 U_\mu(x)\sigma_3) ~~{\rm maximum}
\ee \vspace{-0.2cm}

\noindent where the maximum must be found over the set of all gauge transformations.
Not knowing the value of $\beta_{\rm eff}$,
we measure separately $H_1(\beta)$ and $H_2(\beta)$.
In fact, we can try to infer $\beta_{\rm eff}$ from the requirement that
Eq.(\ref{eq:hel_mod_non_abelian}) represents an order parameter.
The value $\beta_{\rm eff}=1.0$ around the transition region works well, as shown  Fig.\ref{fig:hm_Mag_b_Beff1}.

The similarity with Fig.\ref{fig:hm_rew_8p4_10p4_12p4_14p4} is remarkable, but finite size
effects here are more suggestive of a second order phase transition, as they should be.

This behavior can be considered as a strong support of {\em both} assumptions (1) and (2).
Our measurements of the same observable Eq.(11) after fixing to another gauge do not show a
similar behavior. This presumably singles out the Maximal Abelian Gauge as yielding an effective action
particularly local and close to the Wilson action. Fixing to another gauge,
the effective action contains sizeable
additional terms, and the expression for the helicity modulus Eq.(\ref{eq:hm_def}) differs
appreciably from Eq.(\ref{eq:hel_mod_non_abelian}).

\end{document}